\documentclass[twocolumn,showpacs,aps,prl]{revtex4}

\newcommand{\bk}{{\bf k}}
\newcommand{\ba}{{\bf a}}
\newcommand{\bb}{{\bf b}}

\newcommand{\bm}{{\bf M}}

\newcommand{\be}{{\bf e}}

\newcommand{\beq}{\begin{eqnarray}}
\newcommand{\eeq}{\end{eqnarray}}
\newcommand{\beqq}{\begin{eqnarray*}}
\newcommand{\eeqq}{\end{eqnarray*}}

\usepackage{graphicx}
\usepackage{dcolumn}
\usepackage{bm}
\usepackage{color}

\begin{document}

\begin{titlepage}

\title{Tunable Weyl fermions and Fermi arcs in \\ magnetized topological crystalline insulators}

\author{Junwei Liu$^1$, Chen Fang$^{2}$, and Liang Fu$^{1}$}
\affiliation{$^1$Department of physics, Massachusetts Institute of Technology, Cambridge, MA 02139, USA}
\affiliation{$^2$Beijing National Laboratory for Condensed Matter Physics, and Institute of Physics, Chinese Academy of Sciences, Beijing 100190, China}

\date{\today}

\begin{abstract}
Based on $k\cdot p$ analysis and realistic tight-binding calculations, we find that time-reversal-breaking Weyl semimetals can be realized in magnetically-doped (Mn, Eu, Cr etc.) Sn$_{1-x}$Pb$_x$(Te,Se) class of topological crystalline insulators. All the Weyl points are well separated in momentum space and possess nearly the same energy due to high crystalline symmetry.
Moreover, both the Weyl points and Fermi arcs are highly tunable by varying Pb/Sn composition, pressure, magnetization, temperature, surface potential etc., opening up the possibility of manipulating Weyl points and rewiring the Fermi arcs.
\end{abstract}


\maketitle

\draft

\vspace{2mm}

\end{titlepage}

Weyl semimetals have doubly degenerate band crossings at isolated points (Weyl points) in Brillouin zone (BZ), with linear energy-momentum relation at low energy similar to Weyl fermions in high-energy physics\cite{Weyl1929}. These Weyl points in the band structure are the source or sink of Berry curvature in momentum space, and their presence gives rise to exotic phenomena such as topological Fermi arcs on the surfaces\cite{Wan2011}, the chiral anomaly\cite{Nielsen1983,Son2013,Liu2013} and unusual quantum oscillations\cite{Potter2014}, thus attracting tremendous interest\cite{Witten2015}.

Weyl semimetals can be realized in two classes of materials, with broken inversion or time-reversal symmetry respectively\cite{Young2012}. Recently, the inversion-breaking Weyl semimetal phase has been predicted and observed in the TaAs class of materials\cite{Weng2015,Huang2015,Xu2015a,Xu2015b,Lv2015,Lv2015a,Yang2015}, and similar Weyl points have been found in photonic crystals\cite{Lu2013,Lu2014,Lu2015}. On the other hand, although long sought after\cite{Wan2011, Burkov2011,Burkov2011a,Xu2011,Fang2012,William2012,HeonJung2013,Daniel2014,Tena2015,Borisenko2015,chang2016}, a time-reversal-breaking Weyl semimetal phase has yet be definitively found in real materials. The challenge is to find strong exchange coupling between magnetic order and itinerant electrons in a low (ideally zero) carrier density system.
Among other things, magnetic Weyl semimetals are predicted to exhibit a unique anomalous Hall effect resulting from a nonzero net Berry curvature in momentum space\cite{Ran2011}.

The recently discovered Sn$_{1-x}$Pb$_x$(Te,Se) class of topological crystalline insulators (TCIs)\cite{Fu2011,Hsieh2012,Dziawa2012,Tanaka2012,Xu2012} offer an ideal material base for realizing magnetic Weyl semimetals. These materials are IV-VI semiconductors with a simple rocksalt structure. The nontrivial topology originates from the band inversion at four $L$ points in the BZ\cite{Hsieh2012}, which has been experimentally demonstrated by varying the Pb/Sn composition\cite{Tanaka2012,Xu2012,yan2014,zeljkovic2015dirac} or the temperature\cite{Dziawa2012}. At the gap closing point, the low-energy band structure consists of a massless Dirac fermion at each $L$ valley, which is equivalent to a pair of Weyl fermions with opposite chiralities. This Weyl pair can in principle be separated in momentum space by spin splitting in a Zeeman field\cite{Cho2011}, resulting in the desired magnetic Weyl semimetal. In this regard, it was established decades ago that
very diluted magnetic doping can induce ferromagnetic order in SnTe and other IV-VI semiconductors\cite{Story1986, bauer1986}. Numerous following experimental and theoretical studies have shown that the exchange coupling between magnetic moments and itinerant electrons is  strong and the resulting spin splitting is relatively large\cite{Gorska1988,Pascher1989,Rothlein1990, Gazka1991, Hofmann1992, Bauer1992, Dietl1994, Jonge1990,story1992,Eggenkamp1993,Eggenkamp1995,Geist1997}.
These unique properties make IV-VI semiconductors, especially those near the topological phase transition, attractive in pursuing magnetic Weyl semimetals.

In this work, based on $k\cdot p$ theory and realistic tight-binding (TB) calculations, we report the theoretical discovery of the Weyl semimetal phase in magnetized IV-VI semiconductors Sn$_{1-x}$Pb$_x$(Te,Se). The pair of Weyl points near each $L$ point are widely separated in momentum space by as large as 0.05 $\rm\AA^{-1}$, and all Weyl points have nearly the same energy due to the high crystal symmetry. In addition, by tuning Pb/Sn composition or applying magnetic field, external pressure or strain, one can create, manipulate and annihilate Weyl points in momentum space, and achieve a sequence of topological phase transitions between Weyl semimetals and ferromagnetic insulators in both trivial and TCI phase.

\begin{figure}[tbp]
\includegraphics[width=8cm]{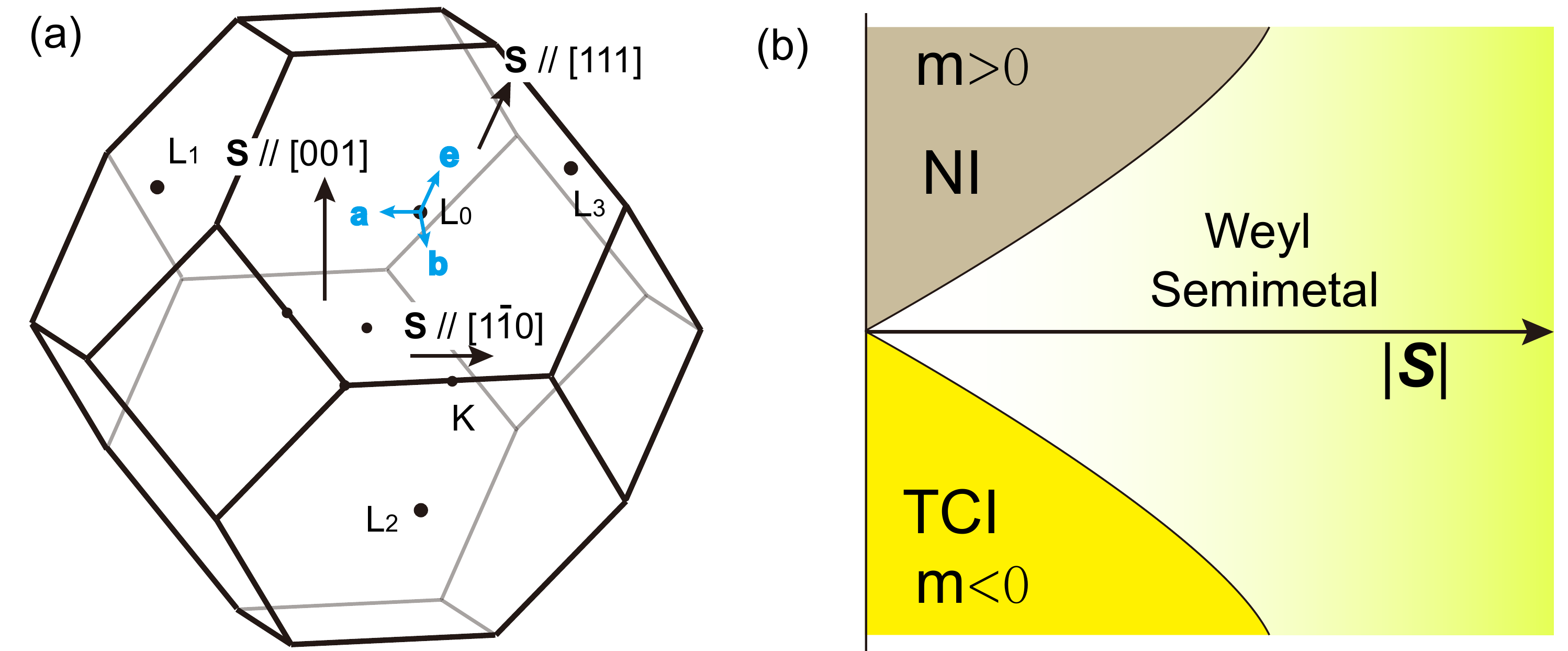}
\caption{(a) Bulk Brillouin zone with four equivalent $L$ points and the possible directions of magnetization $\bm S$.
(b) schematic show of all the phases: normal insulator (NI), topological crystalline insulator (TCI) and Weyl semimetal, varying with the mass $m$ and magnetization $|\bm S|$. 
}
\end{figure}

Pb$_x$Sn$_{1-x}$(Te,Se) are narrow-gap semiconductors with band gap minima at $L$ points. The low-energy properties near $L_{j=1,2,3.4}$ can be well described by the massive Dirac fermions and the $k\cdot p$ Hamiltonian is \cite{Mitchell1966}
\beq
H^0_j(\bk) =m \sigma_z + v_\perp  (\bk \cdot \ba_j \sigma_x s_y - \bk \cdot \bb_j \sigma_x s_x) + v_\parallel \bk \cdot \be_j \sigma_y \nonumber
\eeq
where $\{\ba_j, \bb_j, \be_j \}$ is a local coordinate system at $L_j$ with $\be_j$ along $\Gamma L_j$ direction and $\ba_j$ perpendicular the plane containing $\Gamma$ and $L_j$ (see Fig. 1), and $\bk$ is the momentum relative to $L_j$. $\sigma_z=\pm 1$ represents the $p$-orbital on the cation and anion respectively, and $s_z$ labels the total angular momentum $j_z=\pm 1/2$ along $\be_j$ direction. Based on symmetry analysis of point group $D_{3d}$, one can deduce the general form of exchange couple between conducting electrons and local magnetic moment $\bm S$\cite{Bauer1992,Dietl1994}
\beq
V_j&=& (g_\parallel I + g_\parallel' \sigma_z) \otimes (S_x s_x + S_y s_y)  \nonumber \\
&+&  (g_\perp I + g_\perp' \sigma_z ) \otimes S_z s_z \nonumber
\eeq
where $g_\parallel, g_\parallel', g_\perp$ and $g_\perp'$ are the exchange coupling parameters, and $S_{x,y,z}$ are the components of magnetization along $\ba_j$, $\bb_j$ and $\be_j$. We can set $A=g_\perp+g_\perp', B=g_\perp-g_\perp', a_1=g_\parallel+g_\parallel', b_1=g_\parallel-g_\parallel'$, so that $A$ and $a_1$ ($B$ and $b_1$) represent the exchange coupling between conduction bands (valence bands) and the out-of-plane and in-plane components of the magnetic moments, respectively. Since there is continuous rotation symmetry $C_{\infty z}$ in $H^0_j(\bk)$ to the linear order in $k$, we can safely assume $S_y=0$ without loss of generality. One should be reminded that the $k\cdot{p}$ model to the linear order in $k$ has more symmetries than the real system: the little group at $L$ is enhanced from $D_{3d}$ to $D_{\infty{h}}$, generated by rotation along $\be$, $C_n$, a vertical mirror plane, $M_x (\ba \rightarrow -\ba)$and a mirror plane that flips $k_e$, $M_z (\be \rightarrow -\be)$. This entails, as we will later see, that some band crossings in this effective model are not protected when higher-order terms are included.

In the presence of ferromagnetic ordering, both the conduction and the valence bands become spin-split and the splitting increases with the exchange field (or the magnetization) as shown in Fig. 2, and the top valence band and the bottom conduction band approaches each other, until the gap closes at a critical point
\beq
m=m_c\equiv\frac{1}{2}\left(\sqrt{A^2 S_z^2 + a_1^2 S_y^2}+\sqrt{B^2 S_z^2 + b_1^2 S_y^2}\right)\nonumber
\eeq
If we further increase the exchange coupling, a band inversion between the top valence band (TVB) and bottom conduction band (BCB) occurs.
It is well known that two non-degenerate bands cannot fully anti-cross each other in a 3D BZ after band inversion, but cross each other at an even number of discrete Weyl points in general\cite{Murakami2007}.

Below, we discuss two special cases where the magnetization is along some high-symmetry directions.
When $\bm S\parallel[111]$, the rotation symmetry $C_z$ and the mirror reflection $M_z$ are preserved. If $AB>0$, we find TVB and BCB transform differently under $C_z$ rotation symmetry but identically under $M_z$ mirror symmetry. Since two bands can only cross when they have different quantum numbers, we know that the crossing points in $\ba\bb$ plane will become anti-crossing and the gap opens, while the crossing along $\be$ persists to form a pair of Weyl points (see the upper row of Fig. 2). Instead, for $AB<0$, the crossing bands have the same $C_z$ eigenvalue but different $M_z$ eigenvalues, resulting in a nodal line crossing on the $ab$-plane (see the lower row of Fig. 2). It should be noted, however, that the mirror symmetry $M_z$ is an artifact of the lineared $k \cdot p$ theory, instead of a crystal symmetry in a real rocksalt structure. Therefore, the nodal line on $ab$-plane will generically degrade into two Weyl points.
Next we discuss the case where $\bm S\parallel\ba$. Similarly, there could either be two Weyl points along $\ba$ ($a_1 b_1<0$) or a nodal line in the $\bb\be$-plane ($a_1 b_1>0$) protected by $M_x$ mirror symmetry. Different from the previous case is that $M_x$ is an exact symmetry of the rocksalt structure, and hence the nodal line in this case is robust against all high-order terms.
For a generic magnetization not along any high-symmetry line or plane, the band crossing points are two Weyl points. Firstly we ignore in-plane magnetic moment, and we should get a pair of Weyl points along $\be$ or nodal lines in $\ba\bb$-plane; then, we take into account of the effect of in-plane magnetic moment. Based on the mirror symmetry, it will only shift the location of Weyl points along the in-plane magnetic moment direction or change the nodal line into a pair of Weyl points located in the plane spanned by $\be$ and $\bm S$. 

\begin{figure}[tbp]
\includegraphics[width=8cm]{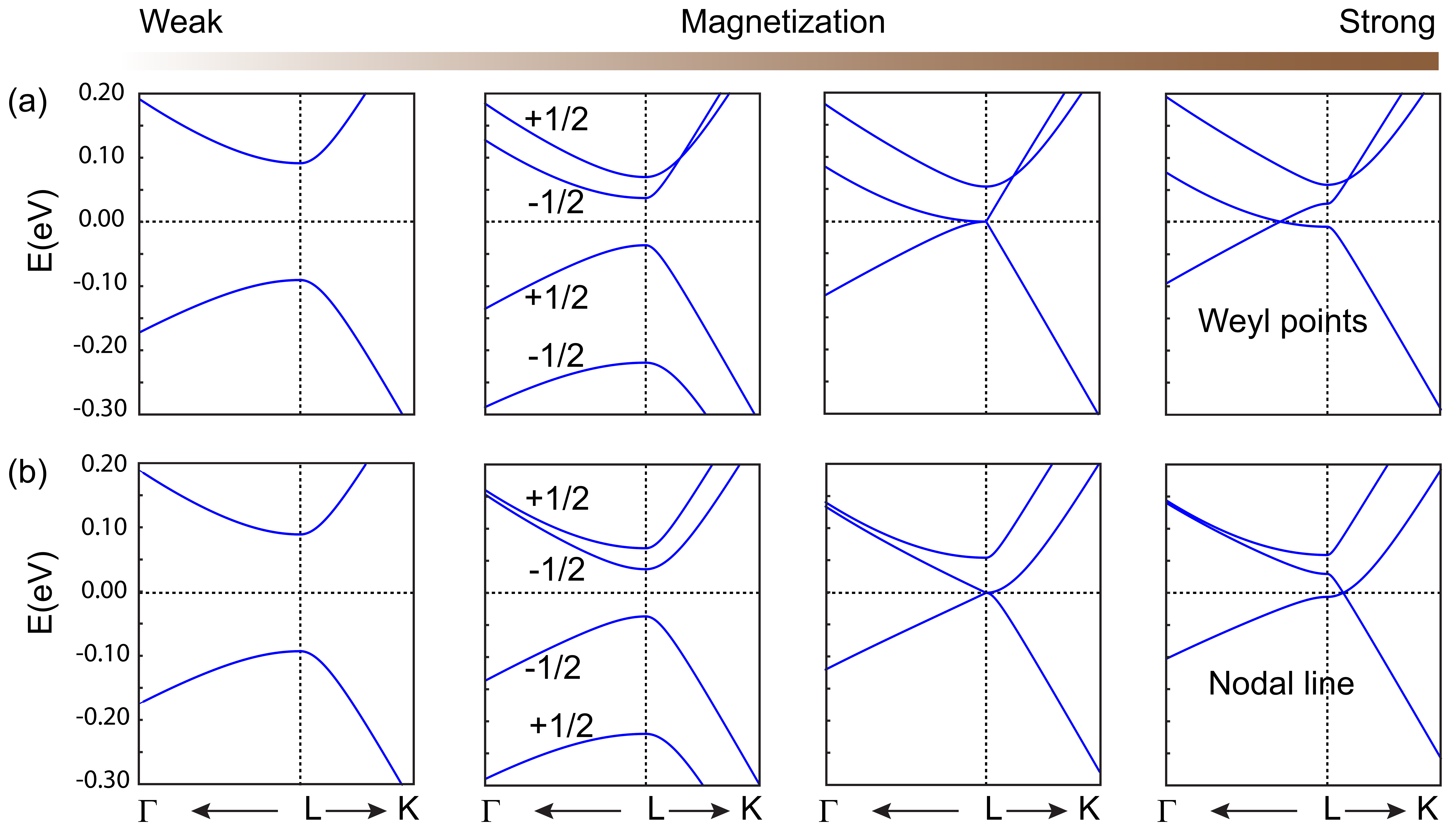}
\caption{Bands evolution with the increase of out-of-plane magnetization along $\Gamma L$ direction. There could be Weyl points along $\Gamma L$ line ($AB>0$) or nodal lines in $k_z=0$ plane ($AB<0$). $\pm 1/2$ labels different eigenvalues of total angular momentum $j_z=\pm 1/2$.
}
\end{figure}

\begin{figure*}[tbp]
\includegraphics[width=17cm]{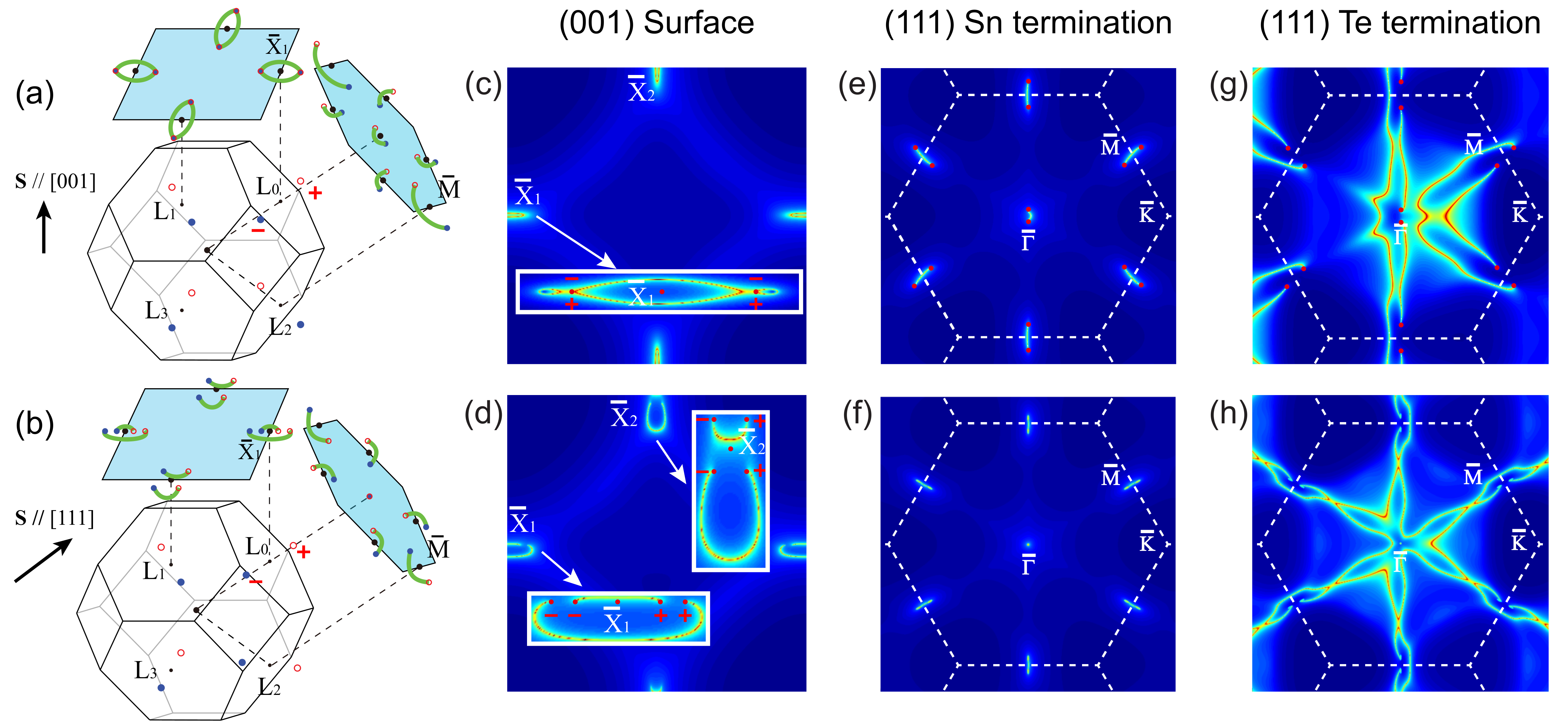}
\caption{Schematic show of Weyl points in the whole BZ and Fermi arcs on (001) and (111) surface with magnetization $\bm S$ along [001] direction (a) and [111] direction (b). Red circle and blue dot represent the Weyl points carrying $+$ and $-$ Chern number. (c-h) are the Fermi arcs directly calculated based on realistic TB model based the previous experimental parameters.
}
\end{figure*}

Now we turn to the analysis for all the four $L$ points in the whole BZ. Since these materials have rock-salt structure with high cubic symmetry, the easy magnetic axes could be $\langle111\rangle$\cite{Eggenkamp1993, Eggenkamp1995}, $\langle110\rangle$ or $\langle001\rangle$. As shown in Fig. 1 and Fig. 3, if the magnetization is along [001] direction, all the four $L$ points are equivalent, related by the rotation and mirror symmetry, setting all the Weyl points at the same energy and located in the $\Gamma L_0 L_2$ or $\Gamma L_1 L_3$ plane based on the previous analysis;
if the magnetization is along [111] direction, the four $L$ points will be divided into two groups: $L_0$ and $L_{1,2,3}$. Near $L_0$, there is a pair of Weyl points along $\Gamma L_0$ direction, while near $L_{1,2,3}$, there is a pair of Weyl points in the $\Gamma L_0 L_{1,2,3}$ plane; if the magnetization is along $[1\bar10]$, the four $L$ points will also be divided into two groups, $L_{0,2}$ and $L_{1,3}$. More interesting, if $a_1 b_1>0$, we simultaneously obtain two nodal lines around $L_{0,2}$ points in $\Gamma L_0 L_2$ plane and two pairs of Weyl points around $L_{1,3}$ points.

Based on the $k\cdot p$ analysis above, we obtain a schematic phase diagram as shown in Fig. 1(b). Without exchange field, the materials could be a normal insulator ($m>0$) or a topological crystalline insulator ($m<0$). At the phase transition point $m=0$, there will be Dirac points at $L$ points, and any finite exchange field will generically separate the Dirac points into pairs of Weyl points with opposite Chern number and thus drive the materials into Weyl semimetal phase. For finite $m$, either inverted or not, when the exchange coupling is sufficiently strong to make $m_c>|m|$, the Weyl semimetal phase is {\it realized}. It was well known that the mass term can be well tuned experimentally by varying temperature\cite{Dziawa2012}, changing the Pb/Sn composition\cite{Tanaka2012,Xu2012,yan2014,zeljkovic2015dirac}, or under internal distortion\cite{Okada2013} or external strain\cite{Qian2014}, while magnetization, both direction and magnitude, can be tuned by an external magnetic field\cite{Eggenkamp1993,Eggenkamp1995,Brodowska2006}. All those properties make these materials have high tunability and thus potentially suitable to study the creation and annihilation of Weyl points.

One of the most interesting properties of Weyl points is that there will be the presence of Fermi arcs ending at the projections of two Weyl points with opposite Chern number. The Fermi arcs are the energy contour of the helicoid Riemann surface states\cite{fang2015}. We can deduce the Fermi arcs based on the locations of Weyl points in the bulk BZ from the bulk-boundary correspondence as shown in the Fig. 3.
If magnetization is along [001] direction, all the four $L$ points can be related by rotation and mirror symmetry, and hence there is a strong confinement for the four pairs of Weyl points. For example, $L_0$ and $L_2$ are related by reflection symmetry $Z \rightarrow -Z$, and therefore the two Weyl points that are projected to the same momentum on (001) surface BZ must carry opposite Chern number as shown in the Fig. 3(a). In this case, there will be no open Fermi arcs on (001) surface, but only closed Fermi loops compose of two Fermi arcs. All these loops can, in principle, shrink to discrete points after adjusting the surface potential; for the (111) surface, all the four pairs of Weyl points are projected to different surface momenta, and there will be robust gapless surface states (Fermi arcs) against any local perturbations.
If the magnetization is along [111] direction, for (001) surface, $\bar X_1$ and $\bar X_2$ are no longer equivalent, thus all the Weyl points are projected to different surface momenta, leading to four different Fermi arcs as shown in Fig. 3(b). However, for (111) surface, the pair of Weyl points around $L_0$ are projected to the same $\bar\Gamma$ point, and there will be no topological protected surface states, but other pairs of Weyl points around $L_{1,2,3}$ points will be projected to different surface momenta along $\bar \Gamma \bar M$ and connected by the nontrivial Fermi arcs.

To further study the material realization and detailed Fermi arcs in the surface BZ, we employed the realistic TB calculations\cite{Lent1986} by adding additional exchange coupling in the $p$ orbital fitted to previous experimental and theoretical studies\cite{Gorska1988,Pascher1989, Rothlein1990, Gazka1991, Hofmann1992, Bauer1992, Dietl1994}.
Here, we take Sn$_{1-x}$Pb$_x$Te as examples, which could be a Dirac semimetal at topological phase transition critical point $x\approx0.65$ .
This TB model was widely used in the past\cite{Lent1986,liu2014spin,Liu2015}, well reproduced experimental bulk electronic structure\cite{Wojek2013,Safaei2013}, and predicted nontrivial topological surface states\cite{Junwei2013} recently confirmed by experiment\cite{yan2014}.
In all previous experimentally studied materials\cite{Bauer1992}, $A$ and $B$ always have the same sign, thus we will always got four pairs of Weyl points. In addition, the distance between a pair of Weyl points in momentum space can reach as large as 0.05 $\rm\AA^{-1}$, within the ARPES resolution. To calculate the Fermi arcs, we employed the recursive Green function method\cite{Sancho1985} to simulate the semi-infinite slab, and the corresponding results are shown in Fig. 3. It is clear that the calculated Fermi arcs are consistent with the $k\cdot p$ analysis. If magnetization is along [001] direction, it is indeed that only (111) surface exhibits Fermi arcs and for [111] magnetization, both (001) and (111) surface states have Fermi arcs. Moreover, the whole Fermi surface is from the non-trivial gapless surface states, which implies that the material is indeed a clean Weyl semimetal with Fermi energy at or very close to the energy of the Weyl points.

\begin{figure}[tbp]
\includegraphics[width=8.5cm]{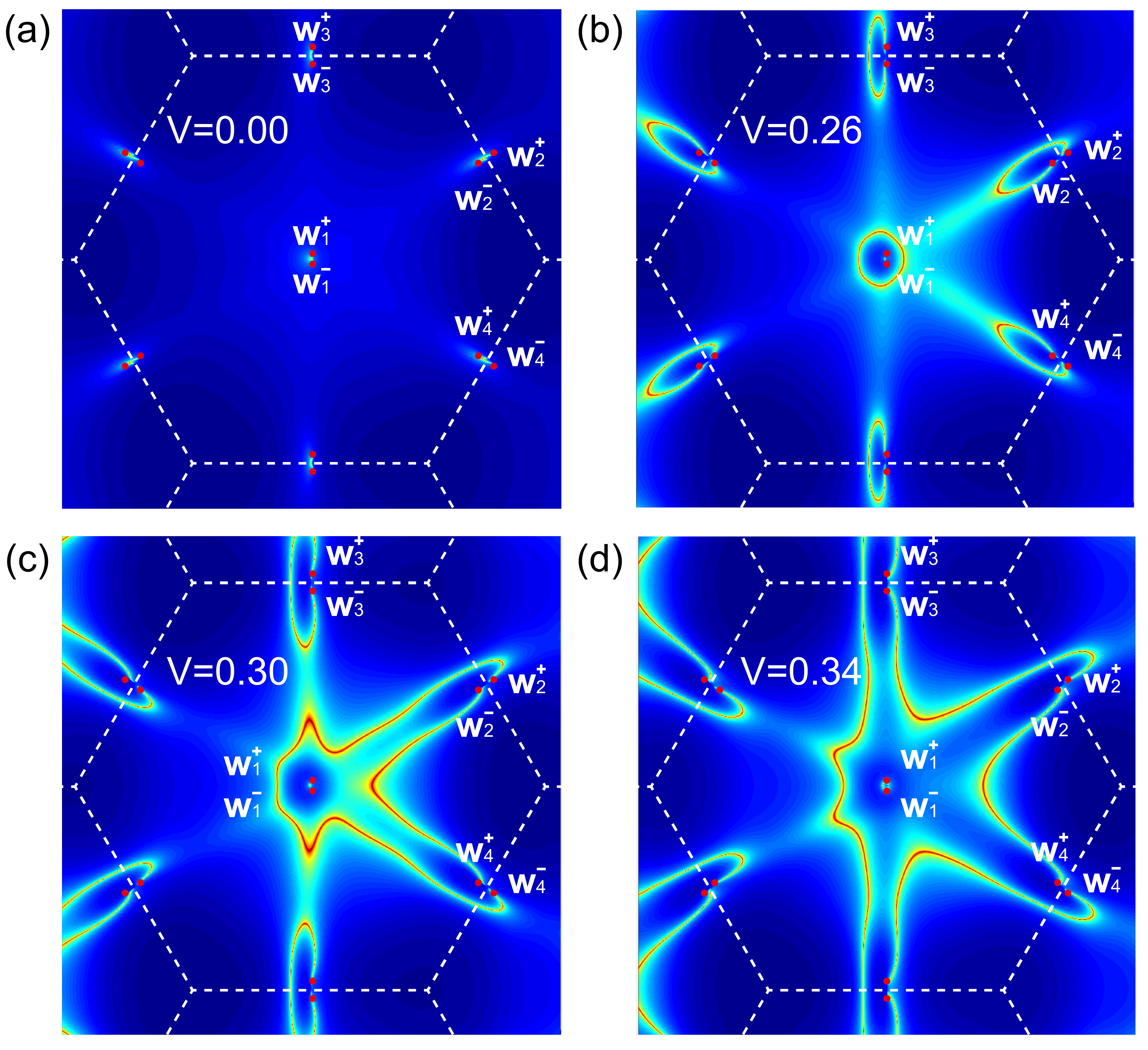}
\caption{The evolution of Fermi arcs varying with the local surface potentials from $V=0.0$ to $V=0.34$. Without potential, each closest pairs of Weyl points $W_{i}^+$ and $W_{i}^-$ ($i=1,2,3,4)$ are connected by the Fermi arcs. Under different potentials, the Fermi arcs can be disconnected and rewired between different Weyl points with opposite Chern number.
}
\end{figure}

The topology analysis can only tell us that there will gapless Riemann surface states and unclosed Fermi arcs on certain surfaces, but the details of those states depend crucially on surface conditions in real materials. For example, there are dramatic differences between the Fermi arcs on the Te-terminated and Sn-terminated (111) surface as shown in the Fig. 3. This suggests the possibility of using surface perturbations to disconnect and reconnect the Fermi arcs. As shown in Fig. 4, by adding a local potential on the surface, we can successfully break the original Fermi arcs connecting $W_{i}^+$ and $W_{i}^-$ $(i=1,2,3,4)$, and rewire the Fermi arcs to connect other Weyl points. Specifically, when $V=0.30$, the Fermi arcs between the pairs of $W_{2,4}^+$ and $W_{2,4}^-$ are disconnected, and the new connections between $W_2^\pm$ and $W_4^\mp$ are established. When we further increase $V$ to $0.34$, the connection between $W_3^+$ and $W_3^-$ is broken and new Fermi arcs are formed between $W_3^+$ ($W_3^-$) and $W_4^-$ ($W_2^+$).
Similar rewiring of Fermi arcs can be achieved by tuning the magnetization. 
As far as we know, the disconnection and reconnection of Fermi arcs have never been experimentally observed or even theoretically studied before, and our results may guide further exploration.

In general, the magnetic doping would increase the trivial gap, thus to realize the Weyl points, it is easier to start from TCI phase with inverted band ordering. However, it should be emphasized that the emergence of Weyl points does not rely on whether the bands are inverted or not, but only depends on the gap size and the exchange coupling of magnetization (Fig. 1(b)). Therefore our analysis and conclusion can be safely applied to other IV-VI semiconductors such as GeTe. Remarkably, it was experimentally found Mn/Cr-doped GeTe are very good ferromagnetic materials\cite{Fukuma2003,dziawa2008} and the Curie temperature could be even higher than 180 K\cite{Fukuma2007,Fukuma2008,dietl2010,Hassan2011}. Our analysis strongly suggests that Mn-doped GeTe could be Weyl semimetals under pressure.

In conclusion, we proposed that magnetic Weyl semimetal can be realized in magnetically-doped (Mn, Eu, Cr) Pb$_x$Sn$_{1-x}$(Te,Se) class of TCIs or the similar IV-VI semiconductors such as GeTe. Depending on the direction of magnetic polarization, the (001) and (111) surfaces exhibited rather different Fermi arcs. And those Fermi arcs can be rewired under local surface potential or external magnetic field, to realize the disconnection and reconnection between different Weyl points, which has not been observed in any experiment. All the results are based on the well-studied experimental parameters, and the separation of Weyl points in momentum space is as large as 0.05 $\rm\AA^{-1}$. We, therefore, believe that it is highly possible to realize and detect the magnetic Weyl points in those materials.

Acknowledgements:
We thank G\"{u}nther Bauer, Tomasz Story and Tomasz Dietl  for helpful discussions.
J. Liu was supported by the MRSEC Program of the National Science Foundation under award number DMR-1419807.
C. Fang was supported by the National Thousand-Young-Talents Program of China.
L. Fu was supported by the DOE Office of Basic Energy Sciences, Division of Materials Sciences and Engineering under Award No. de-sc0010526.

\bibliography{Weyl_ref}

\begin{thebibliography}{70}
\expandafter\ifx\csname natexlab\endcsname\relax\def\natexlab#1{#1}\fi
\expandafter\ifx\csname bibnamefont\endcsname\relax
  \def\bibnamefont#1{#1}\fi
\expandafter\ifx\csname bibfnamefont\endcsname\relax
  \def\bibfnamefont#1{#1}\fi
\expandafter\ifx\csname citenamefont\endcsname\relax
  \def\citenamefont#1{#1}\fi
\expandafter\ifx\csname url\endcsname\relax
  \def\url#1{\texttt{#1}}\fi
\expandafter\ifx\csname urlprefix\endcsname\relax\def\urlprefix{URL }\fi
\providecommand{\bibinfo}[2]{#2}
\providecommand{\eprint}[2][]{\url{#2}}

\bibitem[{\citenamefont{Weyl}(1929)}]{Weyl1929}
\bibinfo{author}{\bibfnamefont{H.}~\bibnamefont{Weyl}},
  \bibinfo{journal}{Zeitschrift f{\"u}r Physik} \textbf{\bibinfo{volume}{56}},
  \bibinfo{pages}{330} (\bibinfo{year}{1929}).

\bibitem[{\citenamefont{Wan et~al.}({2011})\citenamefont{Wan, Turner,
  Vishwanath, and Savrasov}}]{Wan2011}
\bibinfo{author}{\bibfnamefont{X.}~\bibnamefont{Wan}},
  \bibinfo{author}{\bibfnamefont{A.~M.} \bibnamefont{Turner}},
  \bibinfo{author}{\bibfnamefont{A.}~\bibnamefont{Vishwanath}},
  \bibnamefont{and} \bibinfo{author}{\bibfnamefont{S.~Y.}
  \bibnamefont{Savrasov}}, \bibinfo{journal}{Phys. Rev. B}
  \textbf{\bibinfo{volume}{{83}}}, \bibinfo{pages}{205101}
  (\bibinfo{year}{{2011}}).

\bibitem[{\citenamefont{Nielsen and Ninomiya}(1983)}]{Nielsen1983}
\bibinfo{author}{\bibfnamefont{H.}~\bibnamefont{Nielsen}} \bibnamefont{and}
  \bibinfo{author}{\bibfnamefont{M.}~\bibnamefont{Ninomiya}},
  \bibinfo{journal}{Physics Letters B} \textbf{\bibinfo{volume}{130}},
  \bibinfo{pages}{389 } (\bibinfo{year}{1983}).

\bibitem[{\citenamefont{Son and Spivak}(2013)}]{Son2013}
\bibinfo{author}{\bibfnamefont{D.~T.} \bibnamefont{Son}} \bibnamefont{and}
  \bibinfo{author}{\bibfnamefont{B.~Z.} \bibnamefont{Spivak}},
  \bibinfo{journal}{Phys. Rev. B} \textbf{\bibinfo{volume}{88}},
  \bibinfo{pages}{104412} (\bibinfo{year}{2013}).

\bibitem[{\citenamefont{Liu et~al.}(2013{\natexlab{a}})\citenamefont{Liu, Ye,
  and Qi}}]{Liu2013}
\bibinfo{author}{\bibfnamefont{C.-X.} \bibnamefont{Liu}},
  \bibinfo{author}{\bibfnamefont{P.}~\bibnamefont{Ye}}, \bibnamefont{and}
  \bibinfo{author}{\bibfnamefont{X.-L.} \bibnamefont{Qi}},
  \bibinfo{journal}{Phys. Rev. B} \textbf{\bibinfo{volume}{87}},
  \bibinfo{pages}{235306} (\bibinfo{year}{2013}{\natexlab{a}}).

\bibitem[{\citenamefont{Potter et~al.}(2014)\citenamefont{Potter, Kimchi, and
  Vishwanath}}]{Potter2014}
\bibinfo{author}{\bibfnamefont{A.~C.} \bibnamefont{Potter}},
  \bibinfo{author}{\bibfnamefont{I.}~\bibnamefont{Kimchi}}, \bibnamefont{and}
  \bibinfo{author}{\bibfnamefont{A.}~\bibnamefont{Vishwanath}},
  \bibinfo{journal}{Nature Communications} \textbf{\bibinfo{volume}{5}}
  (\bibinfo{year}{2014}).

\bibitem[{\citenamefont{Witten}(2015)}]{Witten2015}
\bibinfo{author}{\bibfnamefont{E.}~\bibnamefont{Witten}},
  \bibinfo{journal}{arXiv preprint arXiv:1510.07698}  (\bibinfo{year}{2015}).

\bibitem[{\citenamefont{Young et~al.}(2012)\citenamefont{Young, Zaheer, Teo,
  Kane, Mele, and Rappe}}]{Young2012}
\bibinfo{author}{\bibfnamefont{S.~M.} \bibnamefont{Young}},
  \bibinfo{author}{\bibfnamefont{S.}~\bibnamefont{Zaheer}},
  \bibinfo{author}{\bibfnamefont{J.~C.~Y.} \bibnamefont{Teo}},
  \bibinfo{author}{\bibfnamefont{C.~L.} \bibnamefont{Kane}},
  \bibinfo{author}{\bibfnamefont{E.~J.} \bibnamefont{Mele}}, \bibnamefont{and}
  \bibinfo{author}{\bibfnamefont{A.~M.} \bibnamefont{Rappe}},
  \bibinfo{journal}{Phys. Rev. Lett.} \textbf{\bibinfo{volume}{108}},
  \bibinfo{pages}{140405} (\bibinfo{year}{2012}).

\bibitem[{\citenamefont{Weng et~al.}(2015)\citenamefont{Weng, Fang, Fang,
  Bernevig, and Dai}}]{Weng2015}
\bibinfo{author}{\bibfnamefont{H.}~\bibnamefont{Weng}},
  \bibinfo{author}{\bibfnamefont{C.}~\bibnamefont{Fang}},
  \bibinfo{author}{\bibfnamefont{Z.}~\bibnamefont{Fang}},
  \bibinfo{author}{\bibfnamefont{B.~A.} \bibnamefont{Bernevig}},
  \bibnamefont{and} \bibinfo{author}{\bibfnamefont{X.}~\bibnamefont{Dai}},
  \bibinfo{journal}{Phys. Rev. X} \textbf{\bibinfo{volume}{5}},
  \bibinfo{pages}{011029} (\bibinfo{year}{2015}).

\bibitem[{\citenamefont{Huang et~al.}(2015)\citenamefont{Huang, Xu, Belopolski,
  Lee, Chang, Wang, Alidoust, Bian, Neupane, Bansil et~al.}}]{Huang2015}
\bibinfo{author}{\bibfnamefont{S.-M.} \bibnamefont{Huang}},
  \bibinfo{author}{\bibfnamefont{S.-Y.} \bibnamefont{Xu}},
  \bibinfo{author}{\bibfnamefont{I.}~\bibnamefont{Belopolski}},
  \bibinfo{author}{\bibfnamefont{C.-C.} \bibnamefont{Lee}},
  \bibinfo{author}{\bibfnamefont{G.}~\bibnamefont{Chang}},
  \bibinfo{author}{\bibfnamefont{B.}~\bibnamefont{Wang}},
  \bibinfo{author}{\bibfnamefont{N.}~\bibnamefont{Alidoust}},
  \bibinfo{author}{\bibfnamefont{G.}~\bibnamefont{Bian}},
  \bibinfo{author}{\bibfnamefont{M.}~\bibnamefont{Neupane}},
  \bibinfo{author}{\bibfnamefont{A.}~\bibnamefont{Bansil}},
  \bibnamefont{et~al.}, \bibinfo{journal}{Nature Communications}
  \textbf{\bibinfo{volume}{6}} (\bibinfo{year}{2015}).

\bibitem[{\citenamefont{Xu et~al.}(2015{\natexlab{a}})\citenamefont{Xu,
  Belopolski, Alidoust, Neupane, Zhang, Sankar, Huang, Lee, Chang, Wang
  et~al.}}]{Xu2015a}
\bibinfo{author}{\bibfnamefont{S.-Y.} \bibnamefont{Xu}},
  \bibinfo{author}{\bibfnamefont{I.}~\bibnamefont{Belopolski}},
  \bibinfo{author}{\bibfnamefont{N.}~\bibnamefont{Alidoust}},
  \bibinfo{author}{\bibfnamefont{M.}~\bibnamefont{Neupane}},
  \bibinfo{author}{\bibfnamefont{C.}~\bibnamefont{Zhang}},
  \bibinfo{author}{\bibfnamefont{R.}~\bibnamefont{Sankar}},
  \bibinfo{author}{\bibfnamefont{S.-M.} \bibnamefont{Huang}},
  \bibinfo{author}{\bibfnamefont{C.-C.} \bibnamefont{Lee}},
  \bibinfo{author}{\bibfnamefont{G.}~\bibnamefont{Chang}},
  \bibinfo{author}{\bibfnamefont{B.}~\bibnamefont{Wang}}, \bibnamefont{et~al.},
  \bibinfo{journal}{Science} \textbf{\bibinfo{volume}{349}},
  \bibinfo{pages}{613} (\bibinfo{year}{2015}{\natexlab{a}}).

\bibitem[{\citenamefont{Xu et~al.}(2015{\natexlab{b}})\citenamefont{Xu,
  Alidoust, Belopolski, Yuan, Bian, Chang, Zheng, Strocov, Sanchez, Chang
  et~al.}}]{Xu2015b}
\bibinfo{author}{\bibfnamefont{S.-Y.} \bibnamefont{Xu}},
  \bibinfo{author}{\bibfnamefont{N.}~\bibnamefont{Alidoust}},
  \bibinfo{author}{\bibfnamefont{I.}~\bibnamefont{Belopolski}},
  \bibinfo{author}{\bibfnamefont{Z.}~\bibnamefont{Yuan}},
  \bibinfo{author}{\bibfnamefont{G.}~\bibnamefont{Bian}},
  \bibinfo{author}{\bibfnamefont{T.-R.} \bibnamefont{Chang}},
  \bibinfo{author}{\bibfnamefont{H.}~\bibnamefont{Zheng}},
  \bibinfo{author}{\bibfnamefont{V.~N.} \bibnamefont{Strocov}},
  \bibinfo{author}{\bibfnamefont{D.~S.} \bibnamefont{Sanchez}},
  \bibinfo{author}{\bibfnamefont{G.}~\bibnamefont{Chang}},
  \bibnamefont{et~al.}, \bibinfo{journal}{Nature Physics}
  \textbf{\bibinfo{volume}{11}}, \bibinfo{pages}{748}
  (\bibinfo{year}{2015}{\natexlab{b}}).

\bibitem[{\citenamefont{Lv et~al.}(2015{\natexlab{a}})\citenamefont{Lv, Weng,
  Fu, Wang, Miao, Ma, Richard, Huang, Zhao, Chen et~al.}}]{Lv2015}
\bibinfo{author}{\bibfnamefont{B.~Q.} \bibnamefont{Lv}},
  \bibinfo{author}{\bibfnamefont{H.~M.} \bibnamefont{Weng}},
  \bibinfo{author}{\bibfnamefont{B.~B.} \bibnamefont{Fu}},
  \bibinfo{author}{\bibfnamefont{X.~P.} \bibnamefont{Wang}},
  \bibinfo{author}{\bibfnamefont{H.}~\bibnamefont{Miao}},
  \bibinfo{author}{\bibfnamefont{J.}~\bibnamefont{Ma}},
  \bibinfo{author}{\bibfnamefont{P.}~\bibnamefont{Richard}},
  \bibinfo{author}{\bibfnamefont{X.~C.} \bibnamefont{Huang}},
  \bibinfo{author}{\bibfnamefont{L.~X.} \bibnamefont{Zhao}},
  \bibinfo{author}{\bibfnamefont{G.~F.} \bibnamefont{Chen}},
  \bibnamefont{et~al.}, \bibinfo{journal}{Phys. Rev. X}
  \textbf{\bibinfo{volume}{5}}, \bibinfo{pages}{031013}
  (\bibinfo{year}{2015}{\natexlab{a}}).

\bibitem[{\citenamefont{Lv et~al.}(2015{\natexlab{b}})\citenamefont{Lv, Xu,
  Weng, Ma, Richard, Huang, Zhao, Chen, Matt, Bisti et~al.}}]{Lv2015a}
\bibinfo{author}{\bibfnamefont{B.~Q.} \bibnamefont{Lv}},
  \bibinfo{author}{\bibfnamefont{N.}~\bibnamefont{Xu}},
  \bibinfo{author}{\bibfnamefont{H.~M.} \bibnamefont{Weng}},
  \bibinfo{author}{\bibfnamefont{J.~Z.} \bibnamefont{Ma}},
  \bibinfo{author}{\bibfnamefont{P.}~\bibnamefont{Richard}},
  \bibinfo{author}{\bibfnamefont{X.~C.} \bibnamefont{Huang}},
  \bibinfo{author}{\bibfnamefont{L.~X.} \bibnamefont{Zhao}},
  \bibinfo{author}{\bibfnamefont{G.~F.} \bibnamefont{Chen}},
  \bibinfo{author}{\bibfnamefont{C.~E.} \bibnamefont{Matt}},
  \bibinfo{author}{\bibfnamefont{F.}~\bibnamefont{Bisti}},
  \bibnamefont{et~al.}, \bibinfo{journal}{Nature Physics}
  \textbf{\bibinfo{volume}{11}}, \bibinfo{pages}{724}
  (\bibinfo{year}{2015}{\natexlab{b}}).

\bibitem[{\citenamefont{Yang et~al.}(2015)\citenamefont{Yang, Liu, Sun, Peng,
  Yang, Zhang, Zhou, Zhang, Guo, Rahn et~al.}}]{Yang2015}
\bibinfo{author}{\bibfnamefont{L.~X.} \bibnamefont{Yang}},
  \bibinfo{author}{\bibfnamefont{Z.~K.} \bibnamefont{Liu}},
  \bibinfo{author}{\bibfnamefont{Y.}~\bibnamefont{Sun}},
  \bibinfo{author}{\bibfnamefont{H.}~\bibnamefont{Peng}},
  \bibinfo{author}{\bibfnamefont{H.~F.} \bibnamefont{Yang}},
  \bibinfo{author}{\bibfnamefont{T.}~\bibnamefont{Zhang}},
  \bibinfo{author}{\bibfnamefont{B.}~\bibnamefont{Zhou}},
  \bibinfo{author}{\bibfnamefont{Y.}~\bibnamefont{Zhang}},
  \bibinfo{author}{\bibfnamefont{Y.~F.} \bibnamefont{Guo}},
  \bibinfo{author}{\bibfnamefont{M.}~\bibnamefont{Rahn}}, \bibnamefont{et~al.},
  \bibinfo{journal}{Nature Physics} \textbf{\bibinfo{volume}{11}},
  \bibinfo{pages}{728} (\bibinfo{year}{2015}).

\bibitem[{\citenamefont{Lu et~al.}(2013)\citenamefont{Lu, Fu, Joannopoulos, and
  Soljacic}}]{Lu2013}
\bibinfo{author}{\bibfnamefont{L.}~\bibnamefont{Lu}},
  \bibinfo{author}{\bibfnamefont{L.}~\bibnamefont{Fu}},
  \bibinfo{author}{\bibfnamefont{J.~D.} \bibnamefont{Joannopoulos}},
  \bibnamefont{and} \bibinfo{author}{\bibfnamefont{M.}~\bibnamefont{Soljacic}},
  \bibinfo{journal}{Nature Photonics}  (\bibinfo{year}{2013}).

\bibitem[{\citenamefont{Lu et~al.}(2014)\citenamefont{Lu, Joannopoulos, and
  Solja{\v{c}}i{\'c}}}]{Lu2014}
\bibinfo{author}{\bibfnamefont{L.}~\bibnamefont{Lu}},
  \bibinfo{author}{\bibfnamefont{J.~D.} \bibnamefont{Joannopoulos}},
  \bibnamefont{and}
  \bibinfo{author}{\bibfnamefont{M.}~\bibnamefont{Solja{\v{c}}i{\'c}}},
  \bibinfo{journal}{Nature Photonics} \textbf{\bibinfo{volume}{8}},
  \bibinfo{pages}{821} (\bibinfo{year}{2014}).

\bibitem[{\citenamefont{Lu et~al.}(2015)\citenamefont{Lu, Wang, Ye, Ran, Fu,
  Joannopoulos, and Soljacic}}]{Lu2015}
\bibinfo{author}{\bibfnamefont{L.}~\bibnamefont{Lu}},
  \bibinfo{author}{\bibfnamefont{Z.}~\bibnamefont{Wang}},
  \bibinfo{author}{\bibfnamefont{D.}~\bibnamefont{Ye}},
  \bibinfo{author}{\bibfnamefont{L.}~\bibnamefont{Ran}},
  \bibinfo{author}{\bibfnamefont{L.}~\bibnamefont{Fu}},
  \bibinfo{author}{\bibfnamefont{J.~D.} \bibnamefont{Joannopoulos}},
  \bibnamefont{and} \bibinfo{author}{\bibfnamefont{M.}~\bibnamefont{Soljacic}},
  \bibinfo{journal}{Science} \textbf{\bibinfo{volume}{349}},
  \bibinfo{pages}{622} (\bibinfo{year}{2015}).

\bibitem[{\citenamefont{Burkov et~al.}({2011})\citenamefont{Burkov, Hook, and
  Balents}}]{Burkov2011}
\bibinfo{author}{\bibfnamefont{A.~A.} \bibnamefont{Burkov}},
  \bibinfo{author}{\bibfnamefont{M.~D.} \bibnamefont{Hook}}, \bibnamefont{and}
  \bibinfo{author}{\bibfnamefont{L.}~\bibnamefont{Balents}},
  \bibinfo{journal}{Phys. Rev. B} \textbf{\bibinfo{volume}{{84}}},
  \bibinfo{pages}{235126} (\bibinfo{year}{{2011}}).

\bibitem[{\citenamefont{Burkov and Balents}(2011)}]{Burkov2011a}
\bibinfo{author}{\bibfnamefont{A.~A.} \bibnamefont{Burkov}} \bibnamefont{and}
  \bibinfo{author}{\bibfnamefont{L.}~\bibnamefont{Balents}},
  \bibinfo{journal}{Phys. Rev. Lett.} \textbf{\bibinfo{volume}{107}},
  \bibinfo{pages}{127205} (\bibinfo{year}{2011}).

\bibitem[{\citenamefont{Xu et~al.}(2011)\citenamefont{Xu, Weng, Wang, Dai, and
  Fang}}]{Xu2011}
\bibinfo{author}{\bibfnamefont{G.}~\bibnamefont{Xu}},
  \bibinfo{author}{\bibfnamefont{H.}~\bibnamefont{Weng}},
  \bibinfo{author}{\bibfnamefont{Z.}~\bibnamefont{Wang}},
  \bibinfo{author}{\bibfnamefont{X.}~\bibnamefont{Dai}}, \bibnamefont{and}
  \bibinfo{author}{\bibfnamefont{Z.}~\bibnamefont{Fang}},
  \bibinfo{journal}{Phys. Rev. Lett.} \textbf{\bibinfo{volume}{107}},
  \bibinfo{pages}{186806} (\bibinfo{year}{2011}).

\bibitem[{\citenamefont{Fang et~al.}(2012)\citenamefont{Fang, Gilbert, Dai, and
  Bernevig}}]{Fang2012}
\bibinfo{author}{\bibfnamefont{C.}~\bibnamefont{Fang}},
  \bibinfo{author}{\bibfnamefont{M.~J.} \bibnamefont{Gilbert}},
  \bibinfo{author}{\bibfnamefont{X.}~\bibnamefont{Dai}}, \bibnamefont{and}
  \bibinfo{author}{\bibfnamefont{B.~A.} \bibnamefont{Bernevig}},
  \bibinfo{journal}{Phys. Rev. Lett.} \textbf{\bibinfo{volume}{108}},
  \bibinfo{pages}{266802} (\bibinfo{year}{2012}).

\bibitem[{\citenamefont{Witczak-Krempa and Kim}(2012)}]{William2012}
\bibinfo{author}{\bibfnamefont{W.}~\bibnamefont{Witczak-Krempa}}
  \bibnamefont{and} \bibinfo{author}{\bibfnamefont{Y.~B.} \bibnamefont{Kim}},
  \bibinfo{journal}{Phys. Rev. B} \textbf{\bibinfo{volume}{85}},
  \bibinfo{pages}{045124} (\bibinfo{year}{2012}).

\bibitem[{\citenamefont{Kim et~al.}(2013)\citenamefont{Kim, Kim, Wang, Sasaki,
  Satoh, Ohnishi, Kitaura, Yang, and Li}}]{HeonJung2013}
\bibinfo{author}{\bibfnamefont{H.-J.} \bibnamefont{Kim}},
  \bibinfo{author}{\bibfnamefont{K.-S.} \bibnamefont{Kim}},
  \bibinfo{author}{\bibfnamefont{J.-F.} \bibnamefont{Wang}},
  \bibinfo{author}{\bibfnamefont{M.}~\bibnamefont{Sasaki}},
  \bibinfo{author}{\bibfnamefont{N.}~\bibnamefont{Satoh}},
  \bibinfo{author}{\bibfnamefont{A.}~\bibnamefont{Ohnishi}},
  \bibinfo{author}{\bibfnamefont{M.}~\bibnamefont{Kitaura}},
  \bibinfo{author}{\bibfnamefont{M.}~\bibnamefont{Yang}}, \bibnamefont{and}
  \bibinfo{author}{\bibfnamefont{L.}~\bibnamefont{Li}}, \bibinfo{journal}{Phys.
  Rev. Lett.} \textbf{\bibinfo{volume}{111}}, \bibinfo{pages}{246603}
  (\bibinfo{year}{2013}).

\bibitem[{\citenamefont{Bulmash et~al.}(2014)\citenamefont{Bulmash, Liu, and
  Qi}}]{Daniel2014}
\bibinfo{author}{\bibfnamefont{D.}~\bibnamefont{Bulmash}},
  \bibinfo{author}{\bibfnamefont{C.-X.} \bibnamefont{Liu}}, \bibnamefont{and}
  \bibinfo{author}{\bibfnamefont{X.-L.} \bibnamefont{Qi}},
  \bibinfo{journal}{Phys. Rev. B} \textbf{\bibinfo{volume}{89}},
  \bibinfo{pages}{081106} (\bibinfo{year}{2014}).

\bibitem[{\citenamefont{Dub\ifmmode~\check{c}\else \v{c}\fi{}ek
  et~al.}(2015)\citenamefont{Dub\ifmmode~\check{c}\else \v{c}\fi{}ek, Kennedy,
  Lu, Ketterle, Solja\ifmmode \check{c}\else \v{c}\fi{}i\ifmmode~\acute{c}\else
  \'{c}\fi{}, and Buljan}}]{Tena2015}
\bibinfo{author}{\bibfnamefont{T.}~\bibnamefont{Dub\ifmmode~\check{c}\else
  \v{c}\fi{}ek}}, \bibinfo{author}{\bibfnamefont{C.~J.} \bibnamefont{Kennedy}},
  \bibinfo{author}{\bibfnamefont{L.}~\bibnamefont{Lu}},
  \bibinfo{author}{\bibfnamefont{W.}~\bibnamefont{Ketterle}},
  \bibinfo{author}{\bibfnamefont{M.}~\bibnamefont{Solja\ifmmode \check{c}\else
  \v{c}\fi{}i\ifmmode~\acute{c}\else \'{c}\fi{}}}, \bibnamefont{and}
  \bibinfo{author}{\bibfnamefont{H.}~\bibnamefont{Buljan}},
  \bibinfo{journal}{Phys. Rev. Lett.} \textbf{\bibinfo{volume}{114}},
  \bibinfo{pages}{225301} (\bibinfo{year}{2015}).

\bibitem[{\citenamefont{Borisenko et~al.}(2015)\citenamefont{Borisenko,
  Evtushinsky, Gibson, Yaresko, Kim, Ali, Buechner, Hoesch, and
  Cava}}]{Borisenko2015}
\bibinfo{author}{\bibfnamefont{S.}~\bibnamefont{Borisenko}},
  \bibinfo{author}{\bibfnamefont{D.}~\bibnamefont{Evtushinsky}},
  \bibinfo{author}{\bibfnamefont{Q.}~\bibnamefont{Gibson}},
  \bibinfo{author}{\bibfnamefont{A.}~\bibnamefont{Yaresko}},
  \bibinfo{author}{\bibfnamefont{T.}~\bibnamefont{Kim}},
  \bibinfo{author}{\bibfnamefont{M.~N.} \bibnamefont{Ali}},
  \bibinfo{author}{\bibfnamefont{B.}~\bibnamefont{Buechner}},
  \bibinfo{author}{\bibfnamefont{M.}~\bibnamefont{Hoesch}}, \bibnamefont{and}
  \bibinfo{author}{\bibfnamefont{R.~J.} \bibnamefont{Cava}},
  \bibinfo{journal}{arXiv preprint arXiv:1507.04847}  (\bibinfo{year}{2015}).

\bibitem[{\citenamefont{Chang et~al.}(2016)\citenamefont{Chang, Xu, Zheng,
  Singh, Hsu, Belopolski, Sanchez, Bian, Alidoust, Lin et~al.}}]{chang2016}
\bibinfo{author}{\bibfnamefont{G.}~\bibnamefont{Chang}},
  \bibinfo{author}{\bibfnamefont{S.-Y.} \bibnamefont{Xu}},
  \bibinfo{author}{\bibfnamefont{H.}~\bibnamefont{Zheng}},
  \bibinfo{author}{\bibfnamefont{B.}~\bibnamefont{Singh}},
  \bibinfo{author}{\bibfnamefont{C.-H.} \bibnamefont{Hsu}},
  \bibinfo{author}{\bibfnamefont{I.}~\bibnamefont{Belopolski}},
  \bibinfo{author}{\bibfnamefont{D.~S.} \bibnamefont{Sanchez}},
  \bibinfo{author}{\bibfnamefont{G.}~\bibnamefont{Bian}},
  \bibinfo{author}{\bibfnamefont{N.}~\bibnamefont{Alidoust}},
  \bibinfo{author}{\bibfnamefont{H.}~\bibnamefont{Lin}}, \bibnamefont{et~al.},
  \bibinfo{journal}{arXiv preprint arXiv:1603.01255}  (\bibinfo{year}{2016}).

\bibitem[{\citenamefont{Yang et~al.}(2011)\citenamefont{Yang, Lu, and
  Ran}}]{Ran2011}
\bibinfo{author}{\bibfnamefont{K.-Y.} \bibnamefont{Yang}},
  \bibinfo{author}{\bibfnamefont{Y.-M.} \bibnamefont{Lu}}, \bibnamefont{and}
  \bibinfo{author}{\bibfnamefont{Y.}~\bibnamefont{Ran}},
  \bibinfo{journal}{Phys. Rev. B} \textbf{\bibinfo{volume}{84}},
  \bibinfo{pages}{075129} (\bibinfo{year}{2011}).

\bibitem[{\citenamefont{Fu}(2011)}]{Fu2011}
\bibinfo{author}{\bibfnamefont{L.}~\bibnamefont{Fu}}, \bibinfo{journal}{Phys.
  Rev. Lett.} \textbf{\bibinfo{volume}{106}}, \bibinfo{pages}{106802}
  (\bibinfo{year}{2011}).

\bibitem[{\citenamefont{Hsieh et~al.}(2012)\citenamefont{Hsieh, Lin, Liu, Duan,
  Bansil, and Fu}}]{Hsieh2012}
\bibinfo{author}{\bibfnamefont{T.~H.} \bibnamefont{Hsieh}},
  \bibinfo{author}{\bibfnamefont{H.}~\bibnamefont{Lin}},
  \bibinfo{author}{\bibfnamefont{J.}~\bibnamefont{Liu}},
  \bibinfo{author}{\bibfnamefont{W.}~\bibnamefont{Duan}},
  \bibinfo{author}{\bibfnamefont{A.}~\bibnamefont{Bansil}}, \bibnamefont{and}
  \bibinfo{author}{\bibfnamefont{L.}~\bibnamefont{Fu}},
  \bibinfo{journal}{Nature Communications} \textbf{\bibinfo{volume}{3}},
  \bibinfo{pages}{982} (\bibinfo{year}{2012}).

\bibitem[{\citenamefont{Dziawa et~al.}(2012)\citenamefont{Dziawa, Kowalski,
  Dybko, Buczko, Szczerbakow, Szot, Lusakowska, Balasubramanian, Wojek,
  Bernsten et~al.}}]{Dziawa2012}
\bibinfo{author}{\bibfnamefont{P.}~\bibnamefont{Dziawa}},
  \bibinfo{author}{\bibfnamefont{B.~J.} \bibnamefont{Kowalski}},
  \bibinfo{author}{\bibfnamefont{K.}~\bibnamefont{Dybko}},
  \bibinfo{author}{\bibfnamefont{R.}~\bibnamefont{Buczko}},
  \bibinfo{author}{\bibfnamefont{A.}~\bibnamefont{Szczerbakow}},
  \bibinfo{author}{\bibfnamefont{M.}~\bibnamefont{Szot}},
  \bibinfo{author}{\bibfnamefont{E.}~\bibnamefont{Lusakowska}},
  \bibinfo{author}{\bibfnamefont{T.}~\bibnamefont{Balasubramanian}},
  \bibinfo{author}{\bibfnamefont{B.~M.} \bibnamefont{Wojek}},
  \bibinfo{author}{\bibfnamefont{M.~H.} \bibnamefont{Bernsten}},
  \bibnamefont{et~al.}, \bibinfo{journal}{Nature Materials}
  \textbf{\bibinfo{volume}{11}}, \bibinfo{pages}{1023} (\bibinfo{year}{2012}).

\bibitem[{\citenamefont{Tanaka et~al.}(2012)\citenamefont{Tanaka, Ren, Sato,
  Nakayama, Souma, Takahashi, Segawa, and Ando}}]{Tanaka2012}
\bibinfo{author}{\bibfnamefont{Y.}~\bibnamefont{Tanaka}},
  \bibinfo{author}{\bibfnamefont{Z.}~\bibnamefont{Ren}},
  \bibinfo{author}{\bibfnamefont{T.}~\bibnamefont{Sato}},
  \bibinfo{author}{\bibfnamefont{K.}~\bibnamefont{Nakayama}},
  \bibinfo{author}{\bibfnamefont{S.}~\bibnamefont{Souma}},
  \bibinfo{author}{\bibfnamefont{T.}~\bibnamefont{Takahashi}},
  \bibinfo{author}{\bibfnamefont{K.}~\bibnamefont{Segawa}}, \bibnamefont{and}
  \bibinfo{author}{\bibfnamefont{Y.}~\bibnamefont{Ando}},
  \bibinfo{journal}{Nature Physics} \textbf{\bibinfo{volume}{8}},
  \bibinfo{pages}{800} (\bibinfo{year}{2012}).

\bibitem[{\citenamefont{Xu et~al.}(2012)\citenamefont{Xu, Liu, Alidoust,
  Neupane, Qian, Belopolski, Denlinger, Wang, Lin, Wray et~al.}}]{Xu2012}
\bibinfo{author}{\bibfnamefont{S.-Y.} \bibnamefont{Xu}},
  \bibinfo{author}{\bibfnamefont{C.}~\bibnamefont{Liu}},
  \bibinfo{author}{\bibfnamefont{N.}~\bibnamefont{Alidoust}},
  \bibinfo{author}{\bibfnamefont{M.}~\bibnamefont{Neupane}},
  \bibinfo{author}{\bibfnamefont{D.}~\bibnamefont{Qian}},
  \bibinfo{author}{\bibfnamefont{I.}~\bibnamefont{Belopolski}},
  \bibinfo{author}{\bibfnamefont{J.~D.} \bibnamefont{Denlinger}},
  \bibinfo{author}{\bibfnamefont{Y.~J.} \bibnamefont{Wang}},
  \bibinfo{author}{\bibfnamefont{H.}~\bibnamefont{Lin}},
  \bibinfo{author}{\bibfnamefont{L.~A.} \bibnamefont{Wray}},
  \bibnamefont{et~al.}, \bibinfo{journal}{Nature Communications}
  \textbf{\bibinfo{volume}{3}}, \bibinfo{pages}{1192} (\bibinfo{year}{2012}).

\bibitem[{\citenamefont{Yan et~al.}(2014)\citenamefont{Yan, Liu, Zang, Wang,
  Wang, Wang, Zhang, Wang, Ma, Ji et~al.}}]{yan2014}
\bibinfo{author}{\bibfnamefont{C.}~\bibnamefont{Yan}},
  \bibinfo{author}{\bibfnamefont{J.}~\bibnamefont{Liu}},
  \bibinfo{author}{\bibfnamefont{Y.}~\bibnamefont{Zang}},
  \bibinfo{author}{\bibfnamefont{J.}~\bibnamefont{Wang}},
  \bibinfo{author}{\bibfnamefont{Z.}~\bibnamefont{Wang}},
  \bibinfo{author}{\bibfnamefont{P.}~\bibnamefont{Wang}},
  \bibinfo{author}{\bibfnamefont{Z.-D.} \bibnamefont{Zhang}},
  \bibinfo{author}{\bibfnamefont{L.}~\bibnamefont{Wang}},
  \bibinfo{author}{\bibfnamefont{X.}~\bibnamefont{Ma}},
  \bibinfo{author}{\bibfnamefont{S.}~\bibnamefont{Ji}}, \bibnamefont{et~al.},
  \bibinfo{journal}{Phys. Rev. Lett.} \textbf{\bibinfo{volume}{112}},
  \bibinfo{pages}{186801} (\bibinfo{year}{2014}).

\bibitem[{\citenamefont{Zeljkovic et~al.}(2015)\citenamefont{Zeljkovic, Okada,
  Serbyn, Sankar, Walkup, Zhou, Liu, Chang, Wang, Hasan
  et~al.}}]{zeljkovic2015dirac}
\bibinfo{author}{\bibfnamefont{I.}~\bibnamefont{Zeljkovic}},
  \bibinfo{author}{\bibfnamefont{Y.}~\bibnamefont{Okada}},
  \bibinfo{author}{\bibfnamefont{M.}~\bibnamefont{Serbyn}},
  \bibinfo{author}{\bibfnamefont{R.}~\bibnamefont{Sankar}},
  \bibinfo{author}{\bibfnamefont{D.}~\bibnamefont{Walkup}},
  \bibinfo{author}{\bibfnamefont{W.}~\bibnamefont{Zhou}},
  \bibinfo{author}{\bibfnamefont{J.}~\bibnamefont{Liu}},
  \bibinfo{author}{\bibfnamefont{G.}~\bibnamefont{Chang}},
  \bibinfo{author}{\bibfnamefont{Y.~J.} \bibnamefont{Wang}},
  \bibinfo{author}{\bibfnamefont{M.~Z.} \bibnamefont{Hasan}},
  \bibnamefont{et~al.}, \bibinfo{journal}{Nature materials}
  \textbf{\bibinfo{volume}{14}}, \bibinfo{pages}{318} (\bibinfo{year}{2015}).

\bibitem[{\citenamefont{Cho}(2011)}]{Cho2011}
\bibinfo{author}{\bibfnamefont{G.~Y.} \bibnamefont{Cho}},
  \bibinfo{journal}{arXiv preprint arXiv:1110.1939}  (\bibinfo{year}{2011}).

\bibitem[{\citenamefont{Story et~al.}(1986)\citenamefont{Story,
  Ga\l{}a\ifmmode~\mbox{\c{}}\else \c{}\fi{}zka, Frankel, and
  Wolff}}]{Story1986}
\bibinfo{author}{\bibfnamefont{T.}~\bibnamefont{Story}},
  \bibinfo{author}{\bibfnamefont{R.~R.}
  \bibnamefont{Ga\l{}a\ifmmode~\mbox{\c{}}\else \c{}\fi{}zka}},
  \bibinfo{author}{\bibfnamefont{R.~B.} \bibnamefont{Frankel}},
  \bibnamefont{and} \bibinfo{author}{\bibfnamefont{P.~A.} \bibnamefont{Wolff}},
  \bibinfo{journal}{Phys. Rev. Lett.} \textbf{\bibinfo{volume}{56}},
  \bibinfo{pages}{777} (\bibinfo{year}{1986}).

\bibitem[{\citenamefont{Bauer}(1986)}]{bauer1986}
\bibinfo{author}{\bibfnamefont{G.}~\bibnamefont{Bauer}}, in
  \emph{\bibinfo{booktitle}{MRS Proceedings}} (\bibinfo{organization}{Cambridge
  Univ Press}, \bibinfo{year}{1986}), vol.~\bibinfo{volume}{89}, p.
  \bibinfo{pages}{107}.

\bibitem[{\citenamefont{G\'orska and Anderson}(1988)}]{Gorska1988}
\bibinfo{author}{\bibfnamefont{M.}~\bibnamefont{G\'orska}} \bibnamefont{and}
  \bibinfo{author}{\bibfnamefont{J.~R.} \bibnamefont{Anderson}},
  \bibinfo{journal}{Phys. Rev. B} \textbf{\bibinfo{volume}{38}},
  \bibinfo{pages}{9120} (\bibinfo{year}{1988}).

\bibitem[{\citenamefont{Pascher et~al.}(1989)\citenamefont{Pascher, R\"othlein,
  Bauer, and von Ortenberg}}]{Pascher1989}
\bibinfo{author}{\bibfnamefont{H.}~\bibnamefont{Pascher}},
  \bibinfo{author}{\bibfnamefont{P.}~\bibnamefont{R\"othlein}},
  \bibinfo{author}{\bibfnamefont{G.}~\bibnamefont{Bauer}}, \bibnamefont{and}
  \bibinfo{author}{\bibfnamefont{M.}~\bibnamefont{von Ortenberg}},
  \bibinfo{journal}{Phys. Rev. B} \textbf{\bibinfo{volume}{40}},
  \bibinfo{pages}{10469} (\bibinfo{year}{1989}).

\bibitem[{\citenamefont{Rothlein et~al.}(1990)\citenamefont{Rothlein, Pascher,
  Bauer, and Tacke}}]{Rothlein1990}
\bibinfo{author}{\bibfnamefont{P.}~\bibnamefont{Rothlein}},
  \bibinfo{author}{\bibfnamefont{H.}~\bibnamefont{Pascher}},
  \bibinfo{author}{\bibfnamefont{G.}~\bibnamefont{Bauer}}, \bibnamefont{and}
  \bibinfo{author}{\bibfnamefont{M.}~\bibnamefont{Tacke}},
  \bibinfo{journal}{Semiconductor Science and Technology}
  \textbf{\bibinfo{volume}{5}}, \bibinfo{pages}{S147} (\bibinfo{year}{1990}).

\bibitem[{\citenamefont{Galazka et~al.}(1991)\citenamefont{Galazka, Spalek,
  Lewicki, Crooker, Karczewski, and Story}}]{Gazka1991}
\bibinfo{author}{\bibfnamefont{R.~R.} \bibnamefont{Galazka}},
  \bibinfo{author}{\bibfnamefont{J.}~\bibnamefont{Spalek}},
  \bibinfo{author}{\bibfnamefont{A.}~\bibnamefont{Lewicki}},
  \bibinfo{author}{\bibfnamefont{B.~C.} \bibnamefont{Crooker}},
  \bibinfo{author}{\bibfnamefont{G.}~\bibnamefont{Karczewski}},
  \bibnamefont{and} \bibinfo{author}{\bibfnamefont{T.}~\bibnamefont{Story}},
  \bibinfo{journal}{Phys. Rev. B} \textbf{\bibinfo{volume}{43}},
  \bibinfo{pages}{11093} (\bibinfo{year}{1991}).

\bibitem[{\citenamefont{Hofmann et~al.}(1992)\citenamefont{Hofmann, Fichtel,
  Pascher, Frank, and Bauer}}]{Hofmann1992}
\bibinfo{author}{\bibfnamefont{W.}~\bibnamefont{Hofmann}},
  \bibinfo{author}{\bibfnamefont{U.}~\bibnamefont{Fichtel}},
  \bibinfo{author}{\bibfnamefont{H.}~\bibnamefont{Pascher}},
  \bibinfo{author}{\bibfnamefont{N.}~\bibnamefont{Frank}}, \bibnamefont{and}
  \bibinfo{author}{\bibfnamefont{G.}~\bibnamefont{Bauer}},
  \bibinfo{journal}{Phys. Rev. B} \textbf{\bibinfo{volume}{45}},
  \bibinfo{pages}{8742} (\bibinfo{year}{1992}).

\bibitem[{\citenamefont{Bauer et~al.}(1992)\citenamefont{Bauer, Pascher, and
  Zawadzki}}]{Bauer1992}
\bibinfo{author}{\bibfnamefont{G.}~\bibnamefont{Bauer}},
  \bibinfo{author}{\bibfnamefont{H.}~\bibnamefont{Pascher}}, \bibnamefont{and}
  \bibinfo{author}{\bibfnamefont{W.}~\bibnamefont{Zawadzki}},
  \bibinfo{journal}{Semiconductor Science and Technology}
  \textbf{\bibinfo{volume}{7}}, \bibinfo{pages}{703} (\bibinfo{year}{1992}).

\bibitem[{\citenamefont{Dietl et~al.}(1994)\citenamefont{Dietl,
  \ifmmode~\acute{S}\else \'{S}\fi{}liwa, Bauer, and Pascher}}]{Dietl1994}
\bibinfo{author}{\bibfnamefont{T.}~\bibnamefont{Dietl}},
  \bibinfo{author}{\bibfnamefont{C.}~\bibnamefont{\ifmmode~\acute{S}\else
  \'{S}\fi{}liwa}}, \bibinfo{author}{\bibfnamefont{G.}~\bibnamefont{Bauer}},
  \bibnamefont{and} \bibinfo{author}{\bibfnamefont{H.}~\bibnamefont{Pascher}},
  \bibinfo{journal}{Phys. Rev. B} \textbf{\bibinfo{volume}{49}},
  \bibinfo{pages}{2230} (\bibinfo{year}{1994}).

\bibitem[{\citenamefont{de~Jonge et~al.}(1990)\citenamefont{de~Jonge, Swagten,
  Eltink, and Stoffels}}]{Jonge1990}
\bibinfo{author}{\bibfnamefont{W.~J.~M.} \bibnamefont{de~Jonge}},
  \bibinfo{author}{\bibfnamefont{H.~J.~M.} \bibnamefont{Swagten}},
  \bibinfo{author}{\bibfnamefont{S.~J. E.~A.} \bibnamefont{Eltink}},
  \bibnamefont{and} \bibinfo{author}{\bibfnamefont{N.~M.~J.}
  \bibnamefont{Stoffels}}, \bibinfo{journal}{Semiconductor Science and
  Technology} \textbf{\bibinfo{volume}{5}}, \bibinfo{pages}{S131}
  (\bibinfo{year}{1990}).

\bibitem[{\citenamefont{Story et~al.}(1992)\citenamefont{Story, Grodzicka,
  Witkowska, Gorecka, and Dobrowolski}}]{story1992}
\bibinfo{author}{\bibfnamefont{T.}~\bibnamefont{Story}},
  \bibinfo{author}{\bibfnamefont{E.}~\bibnamefont{Grodzicka}},
  \bibinfo{author}{\bibfnamefont{B.}~\bibnamefont{Witkowska}},
  \bibinfo{author}{\bibfnamefont{J.}~\bibnamefont{Gorecka}}, \bibnamefont{and}
  \bibinfo{author}{\bibfnamefont{W.}~\bibnamefont{Dobrowolski}},
  \bibinfo{journal}{Acta Physica Polonica A} \textbf{\bibinfo{volume}{82}},
  \bibinfo{pages}{879} (\bibinfo{year}{1992}).

\bibitem[{\citenamefont{Eggenkamp et~al.}(1993)\citenamefont{Eggenkamp, Story,
  Sw{\"u}ste, Swagten, and de~Jonge}}]{Eggenkamp1993}
\bibinfo{author}{\bibfnamefont{P.}~\bibnamefont{Eggenkamp}},
  \bibinfo{author}{\bibfnamefont{T.}~\bibnamefont{Story}},
  \bibinfo{author}{\bibfnamefont{C.}~\bibnamefont{Sw{\"u}ste}},
  \bibinfo{author}{\bibfnamefont{H.}~\bibnamefont{Swagten}}, \bibnamefont{and}
  \bibinfo{author}{\bibfnamefont{W.}~\bibnamefont{de~Jonge}},
  \bibinfo{journal}{ACTA PHYSICA POLONICA SERIES A}
  \textbf{\bibinfo{volume}{84}}, \bibinfo{pages}{641} (\bibinfo{year}{1993}).

\bibitem[{\citenamefont{Eggenkamp et~al.}(1995)\citenamefont{Eggenkamp,
  Swagten, Story, and de~Jonge}}]{Eggenkamp1995}
\bibinfo{author}{\bibfnamefont{P.}~\bibnamefont{Eggenkamp}},
  \bibinfo{author}{\bibfnamefont{H.}~\bibnamefont{Swagten}},
  \bibinfo{author}{\bibfnamefont{T.}~\bibnamefont{Story}}, \bibnamefont{and}
  \bibinfo{author}{\bibfnamefont{W.}~\bibnamefont{de~Jonge}},
  \bibinfo{journal}{Journal of Magnetism and Magnetic Materials}
  \textbf{\bibinfo{volume}{140-144, Part 3}}, \bibinfo{pages}{2039 }
  (\bibinfo{year}{1995}).

\bibitem[{\citenamefont{Geist et~al.}(1997)\citenamefont{Geist, Herbst,
  Mej\'{\i}a-Garc\'{\i}a, Pascher, Rupprecht, Ueta, Springholz, Bauer, and
  Tacke}}]{Geist1997}
\bibinfo{author}{\bibfnamefont{F.}~\bibnamefont{Geist}},
  \bibinfo{author}{\bibfnamefont{W.}~\bibnamefont{Herbst}},
  \bibinfo{author}{\bibfnamefont{C.}~\bibnamefont{Mej\'{\i}a-Garc\'{\i}a}},
  \bibinfo{author}{\bibfnamefont{H.}~\bibnamefont{Pascher}},
  \bibinfo{author}{\bibfnamefont{R.}~\bibnamefont{Rupprecht}},
  \bibinfo{author}{\bibfnamefont{Y.}~\bibnamefont{Ueta}},
  \bibinfo{author}{\bibfnamefont{G.}~\bibnamefont{Springholz}},
  \bibinfo{author}{\bibfnamefont{G.}~\bibnamefont{Bauer}}, \bibnamefont{and}
  \bibinfo{author}{\bibfnamefont{M.}~\bibnamefont{Tacke}},
  \bibinfo{journal}{Phys. Rev. B} \textbf{\bibinfo{volume}{56}},
  \bibinfo{pages}{13042} (\bibinfo{year}{1997}).

\bibitem[{\citenamefont{Mitchell and Wallis}(1966)}]{Mitchell1966}
\bibinfo{author}{\bibfnamefont{D.~L.} \bibnamefont{Mitchell}} \bibnamefont{and}
  \bibinfo{author}{\bibfnamefont{R.~F.} \bibnamefont{Wallis}},
  \bibinfo{journal}{Phys. Rev.} \textbf{\bibinfo{volume}{151}},
  \bibinfo{pages}{581} (\bibinfo{year}{1966}).

\bibitem[{\citenamefont{Murakami}(2007)}]{Murakami2007}
\bibinfo{author}{\bibfnamefont{S.}~\bibnamefont{Murakami}},
  \bibinfo{journal}{New Journal of Physics} \textbf{\bibinfo{volume}{9}},
  \bibinfo{pages}{356} (\bibinfo{year}{2007}).

\bibitem[{\citenamefont{Okada et~al.}(2013)\citenamefont{Okada, Serbyn, Lin,
  Walkup, Zhou, Dhital, Neupane, Xu, Wang, Sankar et~al.}}]{Okada2013}
\bibinfo{author}{\bibfnamefont{Y.}~\bibnamefont{Okada}},
  \bibinfo{author}{\bibfnamefont{M.}~\bibnamefont{Serbyn}},
  \bibinfo{author}{\bibfnamefont{H.}~\bibnamefont{Lin}},
  \bibinfo{author}{\bibfnamefont{D.}~\bibnamefont{Walkup}},
  \bibinfo{author}{\bibfnamefont{W.}~\bibnamefont{Zhou}},
  \bibinfo{author}{\bibfnamefont{C.}~\bibnamefont{Dhital}},
  \bibinfo{author}{\bibfnamefont{M.}~\bibnamefont{Neupane}},
  \bibinfo{author}{\bibfnamefont{S.}~\bibnamefont{Xu}},
  \bibinfo{author}{\bibfnamefont{Y.~J.} \bibnamefont{Wang}},
  \bibinfo{author}{\bibfnamefont{R.}~\bibnamefont{Sankar}},
  \bibnamefont{et~al.}, \bibinfo{journal}{Science}
  \textbf{\bibinfo{volume}{341}}, \bibinfo{pages}{1496} (\bibinfo{year}{2013}).

\bibitem[{\citenamefont{Qian et~al.}(2014)\citenamefont{Qian, Fu, and
  Li}}]{Qian2014}
\bibinfo{author}{\bibfnamefont{X.}~\bibnamefont{Qian}},
  \bibinfo{author}{\bibfnamefont{L.}~\bibnamefont{Fu}}, \bibnamefont{and}
  \bibinfo{author}{\bibfnamefont{J.}~\bibnamefont{Li}}, \bibinfo{journal}{Nano
  Research} \textbf{\bibinfo{volume}{8}}, \bibinfo{pages}{967}
  (\bibinfo{year}{2014}).

\bibitem[{\citenamefont{Brodowska et~al.}(2006)\citenamefont{Brodowska,
  Dobrowolski, Arciszewska, Slynko, and Dugaev}}]{Brodowska2006}
\bibinfo{author}{\bibfnamefont{B.}~\bibnamefont{Brodowska}},
  \bibinfo{author}{\bibfnamefont{W.}~\bibnamefont{Dobrowolski}},
  \bibinfo{author}{\bibfnamefont{M.}~\bibnamefont{Arciszewska}},
  \bibinfo{author}{\bibfnamefont{E.}~\bibnamefont{Slynko}}, \bibnamefont{and}
  \bibinfo{author}{\bibfnamefont{V.}~\bibnamefont{Dugaev}},
  \bibinfo{journal}{Journal of Alloys and Compounds}
  \textbf{\bibinfo{volume}{423}}, \bibinfo{pages}{205 } (\bibinfo{year}{2006}).

\bibitem[{\citenamefont{Fang et~al.}(2015)\citenamefont{Fang, Lu, Liu, and
  Fu}}]{fang2015}
\bibinfo{author}{\bibfnamefont{C.}~\bibnamefont{Fang}},
  \bibinfo{author}{\bibfnamefont{L.}~\bibnamefont{Lu}},
  \bibinfo{author}{\bibfnamefont{J.}~\bibnamefont{Liu}}, \bibnamefont{and}
  \bibinfo{author}{\bibfnamefont{L.}~\bibnamefont{Fu}}, \bibinfo{journal}{arXiv
  preprint arXiv:1512.01552}  (\bibinfo{year}{2015}).

\bibitem[{\citenamefont{Lent et~al.}(1986)\citenamefont{Lent, Bowen, Dow,
  Allgaier, Sankey, and Ho}}]{Lent1986}
\bibinfo{author}{\bibfnamefont{C.~S.} \bibnamefont{Lent}},
  \bibinfo{author}{\bibfnamefont{M.~A.} \bibnamefont{Bowen}},
  \bibinfo{author}{\bibfnamefont{J.~D.} \bibnamefont{Dow}},
  \bibinfo{author}{\bibfnamefont{R.~S.} \bibnamefont{Allgaier}},
  \bibinfo{author}{\bibfnamefont{O.~F.} \bibnamefont{Sankey}},
  \bibnamefont{and} \bibinfo{author}{\bibfnamefont{E.~S.} \bibnamefont{Ho}},
  \bibinfo{journal}{Superlattices and Microstructures}
  \textbf{\bibinfo{volume}{2}}, \bibinfo{pages}{491 } (\bibinfo{year}{1986}).

\bibitem[{\citenamefont{Liu et~al.}(2014)\citenamefont{Liu, Hsieh, Wei, Duan,
  Moodera, and Fu}}]{liu2014spin}
\bibinfo{author}{\bibfnamefont{J.}~\bibnamefont{Liu}},
  \bibinfo{author}{\bibfnamefont{T.~H.} \bibnamefont{Hsieh}},
  \bibinfo{author}{\bibfnamefont{P.}~\bibnamefont{Wei}},
  \bibinfo{author}{\bibfnamefont{W.}~\bibnamefont{Duan}},
  \bibinfo{author}{\bibfnamefont{J.}~\bibnamefont{Moodera}}, \bibnamefont{and}
  \bibinfo{author}{\bibfnamefont{L.}~\bibnamefont{Fu}},
  \bibinfo{journal}{Nature materials} \textbf{\bibinfo{volume}{13}},
  \bibinfo{pages}{178} (\bibinfo{year}{2014}).

\bibitem[{\citenamefont{Liu et~al.}(2015)\citenamefont{Liu, Qian, and
  Fu}}]{Liu2015}
\bibinfo{author}{\bibfnamefont{J.}~\bibnamefont{Liu}},
  \bibinfo{author}{\bibfnamefont{X.}~\bibnamefont{Qian}}, \bibnamefont{and}
  \bibinfo{author}{\bibfnamefont{L.}~\bibnamefont{Fu}}, \bibinfo{journal}{Nano
  Letters} \textbf{\bibinfo{volume}{15}}, \bibinfo{pages}{2657}
  (\bibinfo{year}{2015}).

\bibitem[{\citenamefont{Wojek et~al.}(2013)\citenamefont{Wojek, Buczko, Safaei,
  Dziawa, Kowalski, Berntsen, Balasubramanian, Leandersson, Szczerbakow, Kacman
  et~al.}}]{Wojek2013}
\bibinfo{author}{\bibfnamefont{B.~M.} \bibnamefont{Wojek}},
  \bibinfo{author}{\bibfnamefont{R.}~\bibnamefont{Buczko}},
  \bibinfo{author}{\bibfnamefont{S.}~\bibnamefont{Safaei}},
  \bibinfo{author}{\bibfnamefont{P.}~\bibnamefont{Dziawa}},
  \bibinfo{author}{\bibfnamefont{B.~J.} \bibnamefont{Kowalski}},
  \bibinfo{author}{\bibfnamefont{M.~H.} \bibnamefont{Berntsen}},
  \bibinfo{author}{\bibfnamefont{T.}~\bibnamefont{Balasubramanian}},
  \bibinfo{author}{\bibfnamefont{M.}~\bibnamefont{Leandersson}},
  \bibinfo{author}{\bibfnamefont{A.}~\bibnamefont{Szczerbakow}},
  \bibinfo{author}{\bibfnamefont{P.}~\bibnamefont{Kacman}},
  \bibnamefont{et~al.}, \bibinfo{journal}{Phys. Rev. B}
  \textbf{\bibinfo{volume}{87}}, \bibinfo{pages}{115106}
  (\bibinfo{year}{2013}).

\bibitem[{\citenamefont{Safaei et~al.}(2013)\citenamefont{Safaei, Kacman, and
  Buczko}}]{Safaei2013}
\bibinfo{author}{\bibfnamefont{S.}~\bibnamefont{Safaei}},
  \bibinfo{author}{\bibfnamefont{P.}~\bibnamefont{Kacman}}, \bibnamefont{and}
  \bibinfo{author}{\bibfnamefont{R.}~\bibnamefont{Buczko}},
  \bibinfo{journal}{Phys. Rev. B} \textbf{\bibinfo{volume}{88}},
  \bibinfo{pages}{045305} (\bibinfo{year}{2013}).

\bibitem[{\citenamefont{Liu et~al.}(2013{\natexlab{b}})\citenamefont{Liu, Duan,
  and Fu}}]{Junwei2013}
\bibinfo{author}{\bibfnamefont{J.}~\bibnamefont{Liu}},
  \bibinfo{author}{\bibfnamefont{W.}~\bibnamefont{Duan}}, \bibnamefont{and}
  \bibinfo{author}{\bibfnamefont{L.}~\bibnamefont{Fu}}, \bibinfo{journal}{Phys.
  Rev. B} \textbf{\bibinfo{volume}{88}}, \bibinfo{pages}{241303}
  (\bibinfo{year}{2013}{\natexlab{b}}).

\bibitem[{\citenamefont{Sancho et~al.}(1985)\citenamefont{Sancho, Sancho,
  Sancho, and Rubio}}]{Sancho1985}
\bibinfo{author}{\bibfnamefont{M.~P.~L.} \bibnamefont{Sancho}},
  \bibinfo{author}{\bibfnamefont{J.~M.~L.} \bibnamefont{Sancho}},
  \bibinfo{author}{\bibfnamefont{J.~M.~L.} \bibnamefont{Sancho}},
  \bibnamefont{and} \bibinfo{author}{\bibfnamefont{J.}~\bibnamefont{Rubio}},
  \bibinfo{journal}{Journal of Physics F: Metal Physics}
  \textbf{\bibinfo{volume}{15}}, \bibinfo{pages}{851} (\bibinfo{year}{1985}).

\bibitem[{\citenamefont{Fukuma et~al.}(2003)\citenamefont{Fukuma, Asada,
  Nishimura, and Koyanagi}}]{Fukuma2003}
\bibinfo{author}{\bibfnamefont{Y.}~\bibnamefont{Fukuma}},
  \bibinfo{author}{\bibfnamefont{H.}~\bibnamefont{Asada}},
  \bibinfo{author}{\bibfnamefont{N.}~\bibnamefont{Nishimura}},
  \bibnamefont{and} \bibinfo{author}{\bibfnamefont{T.}~\bibnamefont{Koyanagi}},
  \bibinfo{journal}{Journal of Applied Physics} \textbf{\bibinfo{volume}{93}},
  \bibinfo{pages}{4034} (\bibinfo{year}{2003}).

\bibitem[{\citenamefont{Dziawa et~al.}(2008)\citenamefont{Dziawa, Knoff,
  Domukhovski, Domagala, Jakiela, Lusakowska, Osinniy, Swiatek, Taliashvili,
  and Story}}]{dziawa2008}
\bibinfo{author}{\bibfnamefont{P.}~\bibnamefont{Dziawa}},
  \bibinfo{author}{\bibfnamefont{W.}~\bibnamefont{Knoff}},
  \bibinfo{author}{\bibfnamefont{V.}~\bibnamefont{Domukhovski}},
  \bibinfo{author}{\bibfnamefont{J.}~\bibnamefont{Domagala}},
  \bibinfo{author}{\bibfnamefont{R.}~\bibnamefont{Jakiela}},
  \bibinfo{author}{\bibfnamefont{E.}~\bibnamefont{Lusakowska}},
  \bibinfo{author}{\bibfnamefont{V.}~\bibnamefont{Osinniy}},
  \bibinfo{author}{\bibfnamefont{K.}~\bibnamefont{Swiatek}},
  \bibinfo{author}{\bibfnamefont{B.}~\bibnamefont{Taliashvili}},
  \bibnamefont{and} \bibinfo{author}{\bibfnamefont{T.}~\bibnamefont{Story}}, in
  \emph{\bibinfo{booktitle}{Narrow Gap Semiconductors 2007}}
  (\bibinfo{publisher}{Springer}, \bibinfo{year}{2008}), pp.
  \bibinfo{pages}{11--14}.

\bibitem[{\citenamefont{Fukuma et~al.}(2007)\citenamefont{Fukuma, Asada,
  Moritake, Irisa, and Koyanagi}}]{Fukuma2007}
\bibinfo{author}{\bibfnamefont{Y.}~\bibnamefont{Fukuma}},
  \bibinfo{author}{\bibfnamefont{H.}~\bibnamefont{Asada}},
  \bibinfo{author}{\bibfnamefont{N.}~\bibnamefont{Moritake}},
  \bibinfo{author}{\bibfnamefont{T.}~\bibnamefont{Irisa}}, \bibnamefont{and}
  \bibinfo{author}{\bibfnamefont{T.}~\bibnamefont{Koyanagi}},
  \bibinfo{journal}{Applied Physics Letters} \textbf{\bibinfo{volume}{91}},
  \bibinfo{pages}{092501} (\bibinfo{year}{2007}).

\bibitem[{\citenamefont{Fukuma et~al.}(2008)\citenamefont{Fukuma, Asada,
  Miyawaki, Koyanagi, Senba, Goto, and Sato}}]{Fukuma2008}
\bibinfo{author}{\bibfnamefont{Y.}~\bibnamefont{Fukuma}},
  \bibinfo{author}{\bibfnamefont{H.}~\bibnamefont{Asada}},
  \bibinfo{author}{\bibfnamefont{S.}~\bibnamefont{Miyawaki}},
  \bibinfo{author}{\bibfnamefont{T.}~\bibnamefont{Koyanagi}},
  \bibinfo{author}{\bibfnamefont{S.}~\bibnamefont{Senba}},
  \bibinfo{author}{\bibfnamefont{K.}~\bibnamefont{Goto}}, \bibnamefont{and}
  \bibinfo{author}{\bibfnamefont{H.}~\bibnamefont{Sato}},
  \bibinfo{journal}{Applied Physics Letters} \textbf{\bibinfo{volume}{93}},
  \bibinfo{eid}{252502} (\bibinfo{year}{2008}).

\bibitem[{\citenamefont{Dietl}(2010)}]{dietl2010}
\bibinfo{author}{\bibfnamefont{T.}~\bibnamefont{Dietl}},
  \bibinfo{journal}{Nature materials} \textbf{\bibinfo{volume}{9}},
  \bibinfo{pages}{965} (\bibinfo{year}{2010}).

\bibitem[{\citenamefont{Hassan et~al.}(2011)\citenamefont{Hassan, Springholz,
  Lechner, Groiss, Kirchschlager, and Bauer}}]{Hassan2011}
\bibinfo{author}{\bibfnamefont{M.}~\bibnamefont{Hassan}},
  \bibinfo{author}{\bibfnamefont{G.}~\bibnamefont{Springholz}},
  \bibinfo{author}{\bibfnamefont{R.}~\bibnamefont{Lechner}},
  \bibinfo{author}{\bibfnamefont{H.}~\bibnamefont{Groiss}},
  \bibinfo{author}{\bibfnamefont{R.}~\bibnamefont{Kirchschlager}},
  \bibnamefont{and} \bibinfo{author}{\bibfnamefont{G.}~\bibnamefont{Bauer}},
  \bibinfo{journal}{Journal of Crystal Growth} \textbf{\bibinfo{volume}{323}},
  \bibinfo{pages}{363 } (\bibinfo{year}{2011}).

\end{thebibliography}

\clearpage
\begin{appendix}

\end{appendix}

\end{document}